\documentclass[showpacs,amsmath,amssymb,prd]{revtex4}
\usepackage{graphicx}
\usepackage{dcolumn}
\usepackage{bm}
\usepackage{epsfig}

\newcommand{\kst}{K^{*}}

\newcommand{\jpsi}{J/\psi}

\newcommand{\ppp}{\pi^+\pi^- \pi^{0}}
\newcommand{\pip}{\pi^+}

\newcommand{\pin}{\pi^-}
\newcommand{\pio}{\pi^0}

\newcommand{\ks}{K^0_S}

\newcommand{\pp}{\pi^+\pi^-}
\newcommand{\kk}{K^+K^-}

\newcommand{\pppp}{\pi^+\pi^-\pi^+\pi^-}
\newcommand{\ppkk}{\pi^+\pi^-K^+K^-}

\newcommand{\kkkk}{K^+K^-K^+K^-}

\newcommand{\phiphi}{\phi\phi}
\newcommand{\gpppp}{\gamma \pi^+\pi^-\pi^+\pi^-}
\newcommand{\gppkk}{\gamma \pi^+\pi^-K^+K^-}

\newcommand{\gkkkk}{\gamma K^+K^-K^+K^-}

\newcommand{\gksks}{\gamma K^0_s K^0_s}
\newcommand{\gphiphi}{\gamma \phi\phi}
\newcommand{\gff}{\gamma f_{2} f_{2}}

\newcommand{\rr}{\rho^{0}\rho^{0}}
\newcommand{\ff}{f_{2} f_{2}}
\newcommand{\kstkstb}{K^{*}{\overline K}^{*}}
\newcommand{\kstokstob}{K^{*0}{\overline K}^{*0}}
\newcommand{\ppppp}{\pi^0 \pi^+\pi^-\pi^+\pi^-}
\newcommand{\pppppp}{\pi^0\pi^+\pi^-\pi^0\pi^+\pi^-}

\newcommand{\omegpp}{\omega \pi^+\pi^-}

\newcommand{\ww}{\omega \omega}

\newcommand{\etac}{\eta_{c}}

\newcommand{\ar}{\rightarrow}

\newcommand{\mpppp}{M_{\pi^+\pi^-\pi^+\pi^-}}
\newcommand{\mppkk}{M_{\pi^+\pi^-K^+K^-}}

\newcommand{\bfg}{\begin{figure}[htpb]}
\newcommand{\efg}{\end{figure}}
\newcommand{\bitm}{\begin{itemize}}
\newcommand{\eitm}{\end{itemize}}
\newcommand{\bnum}{\begin{enumerate}}
\newcommand{\enum}{\end{enumerate}}
\newcommand{\btbl}{\begin{table}[htp]}
\newcommand{\etbl}{\end{table}}
\newcommand{\btbu}{\begin{tabular}[htp]}
\newcommand{\etbu}{\end{tabular}}
\newcommand{\bcl}{\begin{center}}
\newcommand{\ecl}{\end{center}}

\newcommand{\beq}{\begin{equation}}
\newcommand{\eeq}{\end{equation}}
\newcommand{\beqr}{\begin{eqnarray}}
\newcommand{\eeqr}{\end{eqnarray}}

\usepackage{epsfig}

\begin{document}
\normalsize

\parindent = 0.5 in

\title{\boldmath Experimental study of $\etac $ decays into vector-vector final states}

\author{
M.~Ablikim$^{1}$,              J.~Z.~Bai$^{1}$,               Y.~Ban$^{11}$,
J.~G.~Bian$^{1}$,              X.~Cai$^{1}$,                  H.~F.~Chen$^{16}$,
H.~S.~Chen$^{1}$,              H.~X.~Chen$^{1}$,              J.~C.~Chen$^{1}$,
Jin~Chen$^{1}$
% ...
\\
\vspace{0.2cm}
(BES Collaboration)\\
\vspace{0.2cm}
{\it
$^{1}$ Institute of High Energy Physics, Beijing 100049, People's Republic of China\\
% ...
$^{11}$ Peking University, Beijing 100871, People's Republic of China\\
% ...
$^{16}$ University of Science and Technology of China, Hefei 230026, People's Republic of China\\
% ...
}
}


\begin{abstract}

Using 58 million $J/\psi$ events accumulated with the BES\,II
detector, branching fractions for the $\etac\ar VV$ decays, $\etac \ar
\rho \rho$, $\etac \ar \kstkstb$ , $\etac \ar \ww$, and $\etac \ar
\phiphi$, are measured.  The branching fractions are $Br(\etac \ar
\rho \rho) = (1.25 \pm 0.37 \pm 0.51) \times 10^{-2}$, $Br(\etac \ar
\kstkstb) = (10.4 \pm 2.6\pm 4.3) \times 10^{-3}$, $ Br(\etac \ar \ww)
< 6.3 \times 10^{-3}$ (90 \% CL) , and $Br(\etac \ar \phiphi) = (2.5
\pm 0.5 \pm 0.9) \times 10^{-3}$. The process $\etac \ar \omega\phi$
is also searched for, and the 90\% CL upper limit $Br(\etac \ar
\omega\phi) < 1.3 \times 10^{-3}$ is obtained for this process.

\end{abstract}


\maketitle

\section{Introduction}

 It has been known for many years that $\etac$ decays into
vector-vector (VV) mesons, even though this process is forbidden by
hadron helicity conservation (HHC) assuming
collinear valence quark configurations dominate.  Three different
Bethe-Salpeter wave functions (Gauss, Exponent, Power) have been used
to calculate Br($\etac\ar VV$)~\cite{jiayu}; the calculated
results are lower than the experimental ones from the PDG~\cite{pdg} by
two or three orders of magnitude, as shown in Table
\ref{tablebrpdg}. New theoretical predictions are needed.

{\footnotesize
\begin{table}[htpb]
\caption{Theoretical predictions for $Br(\etac\ar VV)$ compared with
PDG results~\cite{pdg}. Gauss, Exponent, and Power refer to three different
Bethe-Salpeter wave functions  used
to calculate Br($\etac\ar VV$)~\cite{jiayu}. }
\begin{center}
\begin{tabular}{|l|c|c|c|c|}
  \hline Br & Gauss & Exponent & Power & PDG \\ \hline
  $Br(\etac\ar\rho\rho)$ & $2.3\times 10^{-5}$ & $8.7 \times 10^{-5}$
  & $2.8\times 10^{-4}$ & $(2.6\pm0.9) \times 10^{-2}$ \\
  $Br(\etac\ar\kstkstb)$ & $2.8\times 10^{-5}$ & $8.6 \times 10^{-5}$
  & $2.8\times 10^{-4}$ & $(8.5\pm3.1) \times 10^{-3}$ \\
  $Br(\etac\ar\phiphi)$ & $4.2\times 10^{-6}$ & $1.6 \times 10^{-5}$ &
  $5.0\times 10^{-5}$ & $(2.6\pm 0.9) \times 10^{-3}$ \\ \hline
\end{tabular}
\label{tablebrpdg}
\end{center}
\end{table}
}


Branching fractions and reduced branching fractions, $BR_{R}$,
compared with SU(3) predictions~\cite{dm2vv} are shown in Table
\ref{tablebr}. Here the reduced branching fraction is defined as $BR_{R}
= BR/(P_{V}^{3} )$, where $P_{V}$ is the momentum of the vector meson
in the $\etac$ center-of-mass system. The experimental results are not
inconsistent with SU(3) expectations, but due to the large errors
no meaningful investigation of possible SU(3) symmetry breaking
patterns can be performed.

{\footnotesize
\begin{table}[htpb]
\caption{Comparison of PDG Br($\etac\ar VV$) results with predictions
  of SU(3) symmetry. Ratio of $BR_{R}$ is
  defined as $BR_{R}(\etac\ar VV)/BR_{R}(\etac\ar\phiphi)$, and this
  can be compared with the SU(3) expectation, SU(3).}
\begin{center}
\begin{tabular}{|l|c|c||c|c|}
  \hline
     & BR($10^{-3}$) & $BR_{R}$($10^{-3}$) & Ratio of $BR_{R}$ & SU(3) \\ \hline
  $Br(\etac\ar\rho\rho)$ & $26\pm9$  & $12.6\pm4.4$ & $6.3\pm2.3$ & 3 \\
  $Br(\etac\ar\kstkstb)$ & $8.5\pm3.1$  & $5\pm1.7$ & $2.5\pm0.9$ & 4  \\
  $Br(\etac\ar\ww)$ & $<3.1$  & $<1.5$ & $<0.8$ & 1  \\
  $Br(\etac\ar\phiphi)$ & $2.6\pm 0.9$  & $2\pm0.7 $ & $1\pm0.4 $ & 1  \\ \hline
\end{tabular}
\label{tablebr}
\end{center}
\end{table}
}

Exclusive decays of $\etac$ into vector meson pairs have also been
investigated in the framework of the $^3P_{0}$ quark-creation
model~\cite{etacvv_width}. The width of $\etac\ar\rr$ in PDG is
significantly larger than the model prediction.  The branching
fraction of $\etac \ar \omega\phi$ is very important to determine the
mechanism of $\etac\ar VV$ decay~\cite{omegphi}.  In this paper,
$J/\psi\to \gamma \etac$ decays from a sample of 58 million $J/\psi$
decays obtained with the BESII detector are used to determine
$Br(\etac\ar\rho\rho)$, $Br(\etac\ar\kstkstb)$, $Br(\etac\ar\ww)$,
$Br(\etac\ar\phiphi)$, and $Br(\etac \ar \omega\phi)$.


\section{BES Detector}

BES is a conventional solenoidal magnet detector~\cite{bes,bes2}.  A
 12-layer vertex chamber (VTC) surrounding the beam pipe provides
 trigger and trajectory information. A forty-layer main drift chamber
 (MDC), located radially outside the VTC, provides trajectory and
 energy loss ($dE/dx$) information for charged tracks over $85\%$ of
 the total solid angle with a momentum resolution of $\sigma _p/p =
 0.0178 \sqrt{1+p^2}$ ($p$ in $\hbox{\rm~GeV}/c$) and a $dE/dx$
 resolution for hadron tracks of $\sim 8\%$. An array of 48
 scintillation counters surrounding the MDC measures the
 time-of-flight (TOF) of charged tracks with a resolution of $\sim
 200$ ps for hadrons.  Radially outside the TOF system is a 12
 radiation length, lead-gas barrel shower counter (BSC).  This
 measures the energies of electrons and photons over $\sim 80\%$ of
 the total solid angle with an energy resolution of
 $\sigma_E/E=21\%/\sqrt{E}$ ($E$ in~GeV).  Outside the solenoidal
 coil, which provides a 0.4~Tesla magnetic field over the tracking
 volume, is an iron flux return that is instrumented with three double
 layers of proportional counters that identify muons of momentum
 greater than 0.5~GeV/c.

A Geant3 based Monte Carlo, SIMBES~\cite{simbes}, which simulates the
detector response, including interactions of secondary particles in
the detector material, is used in this analysis.  Reasonable agreement
between data and Monte Carlo simulation is observed in various
channels tested, including $e^+e^-\to (\gamma) e^+ e^-$,
$e^+e^-\to(\gamma)\mu\mu$, $J/\psi\to p\bar{p}$, $J/\psi\to\rho\pi$,
and $\psi(2S)\to\pi^+\pi^- J/\psi$, $J/\psi\to l^+ l^-$.  For
$\jpsi\ar\gamma\etac$, decays are generated with an angular
distribution of $1+\cos^{2}\theta$, where $\theta$ is the angle between the
$e^{+}$ and $\etac$ in the laboratory.

\section{\boldmath $\etac \ar \rr$ }
\subsection{Event selection}
First a $\jpsi \ar \gpppp$ sample is selected.  Events are required to
have four good charged tracks and one or more photon candidates. A
good track, reconstructed from hits in the MDC, must be well fitted to
a helix originating from the interaction point; have a polar angle,
$\theta$, with $|\cos\theta| < 0.8$; and a transverse momentum greater
than 60 MeV/c.  TOF and $dE/dx$ information are combined to form
particle identification
probabilities for the pion, kaon, and proton hypotheses. At least three
tracks must be identified as pions.
To reduce the number of spurious low energy photons produced by
secondary hadronic interactions, photon candidates must have a minimum
energy of 50 MeV and be outside a cone with a half-angle of
$15^{\circ}$ around any charged track.

To get improved momentum resolution and to remove backgrounds, events
are kinematically fitted to the $\jpsi \ar \gpppp$ hypothesis, using
all photon candidates. The fit with the highest probability is
selected, and the $\chi^2$ of the four constraint (4C) fit is required
to be less than 20.

For $\jpsi\ar\gpppp$ decay, a major source of background is from
$\jpsi \ar \ppppp$. To remove events containing a $\pi^0$, when there
are multiple photons, $|m_{\gamma 1\gamma
2}-m_{\pi^{0}}|>60{\rm~MeV/c^2}$ is required if
$\overrightarrow{P}_{miss}$ is near the plane of the two photons,
${\gamma 1}$ and ${\gamma 2}$, i.e. $|\widehat{P}_{miss}
\cdot({\hat{r}_{\gamma 1}\times \hat{r}_{\gamma
2}})|<0.15$. Here $\widehat{P}_{miss}$ is the unit vector of the
missing momentum of all charged tracks; $\hat{r}_{\gamma 1}$ and
$\hat{r}_{\gamma 2}$ are unit vectors in the ${\gamma 1}$ and ${\gamma
2}$ directions, respectively; and $m_{\gamma 1\gamma
2}$ is the
invariant mass of ${\gamma 1}$ and ${\gamma2}$.  Two additional
requirements, $\chi^{2}({\jpsi\ar\gpppp})<\chi^{2}({\jpsi\ar\gamma\gamma\pppp})$ and
$P^{2}_{t\gamma}<0.0015\rm~(GeV/c)^{2}$, are used to further remove $\jpsi
\ar \ppppp$ background.  $P_{t\gamma}$ is the transverse momentum of
the $\pppp$ system with respect to the photon.  Finally, the
requirement $|U_{miss}|=|E_{miss}- cP_{miss}|<0.07$ $GeV/c$ is used to
reject events with multiple photons or charged kaons; here,
$E_{miss}$ and $P_{miss}$ are, respectively, the missing energy and
missing momentum calculated using only the charged particles, which
are assumed to be pions.

There are other possible backgrounds such as $\jpsi\ar \omegpp$,
$\jpsi\ar \gksks$, and $\jpsi\ar \gamma K^0_s K^{\pm} \pi^{\mp} $.
The $\jpsi\ar \omegpp$ background is suppressed by the requirement
$|M_{\ppp}-M_{\omega}|>40{\rm~MeV/c^2}$, where the $\pio$ in $\ppppp$ is
associated to the missing momentum and energy, determined using only
the charged tracks. To remove the $\jpsi\ar \gksks$ background,
$|M_{\pp}-M_{\ks}|>25{\rm~MeV/c^2}$ is required for both $\pi^+ \pi^-$
pairs. To remove the background from $\jpsi\ar \gamma K^0_s K^{\pm}
\pi^{\mp}$, events are rejected if $\chi^{2}(\jpsi\ar \gamma
K^{\pm}\pi^{\mp}\pip\pin)<\chi^{2}(\jpsi\ar \gpppp)$ when
$|M_{\pp}-M_{\ks}|<25{\rm~MeV/c^2}$.



\begin{figure}[htbp!]
\begin{center}
\epsfxsize=6cm\epsffile{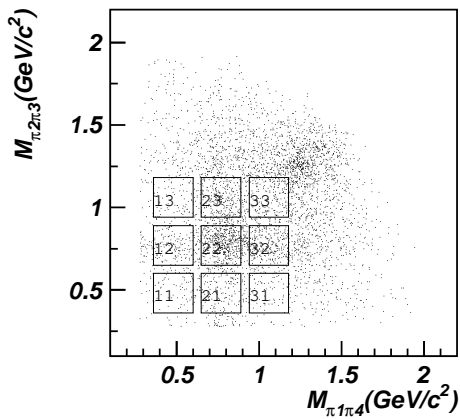}
\vspace*{5pt}
\caption{Scatterplot of  $M_{\pi 2\pi 3}$ versus $M_{\pi 1\pi 4}$.
Also shown are the signal and sideband background boxes used in
this analysis.}
\label{m4piboxrr}
\end{center}
\end{figure}

Figure \ref{m4piboxrr} shows the scatter plot of  $M_{\pi 2\pi 3}$
versus $M_{\pi 1\pi 4}$ for surviving events with $M_{\pppp}\geq
2.8{\rm~GeV}/c^2$, where $\pi 1$ ($\pi 3$) and $\pi 2$ ($\pi 4$) are the
$\pi^+$ and $\pi^-$ with the higher (lower) momentum. There are  clear
signals near ($M_{\rho^0}$, $M_{\rho^0}$) and ($M_{f_{2}(1270)}$,
$M_{f_{2}(1270)}$). The decay $\jpsi\ar\gff$, including
$\etac\ar\ff$, has been studied previously using the $\jpsi$ events
collected by BES II~\cite{gff}.

\begin{figure}[htbp!]
\begin{center}
\epsfxsize=6cm\epsffile{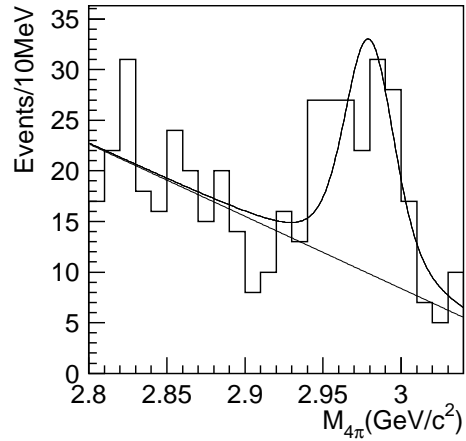}
\vspace*{5pt}
\caption{Fitting $\etac$ in the $\mpppp$ spectrum of events in the
  signal box 22 in Fig.~\ref{m4piboxrr}. The fitted number of $\etac$ events is
  $N^{sig}=106.9\pm18.2$.}
\label{etacmrr}
\end{center}
\end{figure}


The number of $\etac\ar\rr$ events and the corresponding backgrounds
are estimated from the boxes in the scatter plot of $\pp$ versus $\pp$
invariant masses, as shown in Fig. \ref{m4piboxrr}. The signal region
is shown as the square box 22 at (0.77, 0.77)
GeV/c$^{2}$ with a width of 240 MeV/c$^{2}$. The main backgrounds are
$\etac\ar\rho\pp$ and $\etac\ar\pppp$. The $\etac\ar\rho\pp$
background produces the horizontal and vertical bands at
$m_{\rho}$, and the $\etac\ar\pppp$ background produces the uniform
background in the $m_{\pp}$ versus $m_{\pp}$ scatter plot of
Fig. \ref{m4piboxrr}.  These backgrounds are estimated using the
horizontal and vertical sideband background boxes, which are taken 50
MeV/c$^{2}$ away from the signal box and denoted
as 12, 21, 32, and 23, and the two diagonal boxes, denoted as 31 and 13, in
Fig. \ref{m4piboxrr}.

\begin{figure}[htbp!]
\begin{center}
\epsfxsize=6cm\epsffile{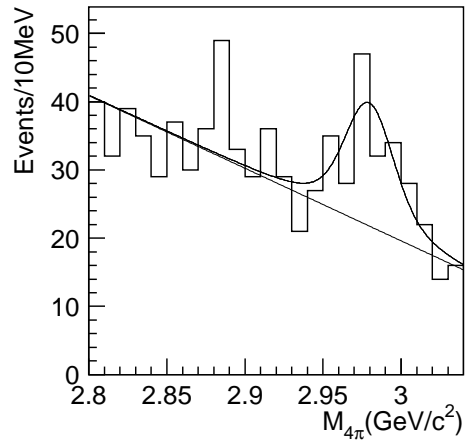}
\vspace*{5pt}
\caption{Fitting $\etac$ in the $\mpppp$ spectrum of four horizontal
and vertical sideband boxes (12,21,32,23). The fitted number of
$\etac$ events is $N^{sid1}=83.3\pm21.5$.}
\label{etacmrr_sideb}
\end{center}
\end{figure}

\begin{figure}[htbp!]
\begin{center}
\epsfxsize=6cm\epsffile{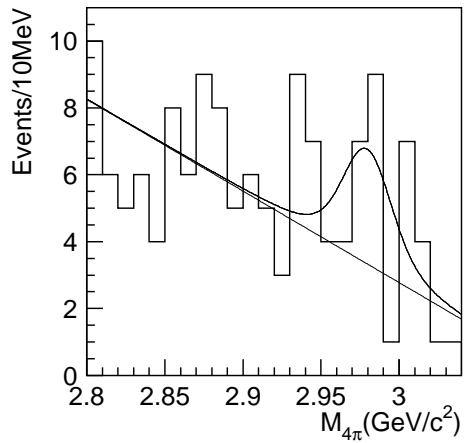}
\vspace*{5pt}
\caption{Fitting $\etac$ in the $\mpppp$
spectrum of two diagonal boxes (31,13). The fitted number of $\etac$
events is $N^{sid2}=15.8\pm8.9$. }
\label{etacmrr_sidec}
\end{center}
\end{figure}


\subsection{Results}

Shown in Figs. \ref{etacmrr}, \ref{etacmrr_sideb}, and
\ref{etacmrr_sidec} are the $\mpppp$ fitting results for $\rr$ signal and
sideband regions.  Binned maximum likelihood fits are performed
with a Breit-Wigner convoluted with a Gaussian resolution function for the
$\etac$ and a first order polynomial for background. The number of
$\etac \ar \rr$ events is obtained with the formula
$N^{obs}=N^{sig}-N^{sid1}/2+N^{sid2}/2$. Here $N^{sig}$, $N^{sid1}$, and
$N^{sid2}$ are the fitted number of $\etac$ events in the $\mpppp$ signal
box and sideband boxes(See Figs. \ref{etacmrr}, \ref{etacmrr_sideb}, \ref{etacmrr_sidec}).  So $N^{obs}
=(106.9\pm18.2)-(83.3\pm21.5)/2+(15.8\pm8.9)/2=73.1\pm21.6$.

The  $\etac \ar \rr$ branching fraction is determined from
\begin{align}
Br(\etac \ar \rr) & = \frac{N^{obs}/\varepsilon}{N_{\jpsi}\cdot Br(\jpsi \ar
  \gamma \etac)} \nonumber \\
 & = (4.15 \pm 1.23) \times 10^{-3}, \nonumber
\end{align}
where $\varepsilon=2.35\%$ is the efficiency, corrected for sideband
  subtraction, determined from Monte Carlo simulation.
  $N_{\jpsi}=57.7\times 10^{6}$ is the total number of $\jpsi$ events
  collected by BES\,II ~\cite{fangss}, and $Br(\jpsi\ar \gamma \etac)
  = 0.013$ is the branching fraction for $\jpsi\ar\gamma
  \etac$~\cite{pdg}. Correcting for charged decay modes, $Br(\etac \ar
  \rho \rho) = (1.25 \pm 0.37) \times 10^{-2}$ is obtained.

\section{\boldmath $\etac \ar \kstkstb$}

  $\jpsi\ar\gamma\etac \ar \gamma K^{*0}{\overline K}^{*0} \ar \gppkk$ is
  studied to determine the $\etac \ar \kstkstb$  branching fraction.

\subsection{Event Selection}

  The selection criteria for charged tracks and photons are the same
as for the $\etac\ar\rr$ analysis. At least one good photon is
required.  TOF and $dE/dx$ information are combined to form particle
identification probabilities for the pion, kaon, and proton
hypotheses for each track, and at least two tracks must be identified as kaons. The
same method as used in the $\etac\ar\rr$ analysis is used to remove
$\pi ^0$ background.

Some additional requirements are applied.  Combined probabilities are
formed from the particle identification probabilities of all tracks
and the probabilities determined from the 4C kinematic fit to the $\jpsi
\ar \gamma \ppkk$ hypothesis for all possible combinations. The fit
with the highest probability is selected, and the $\chi^2$ of the 4C
kinematic fit is required to be less than 10.

To further remove backgrounds containing $\pi ^0$s, two requirements,
$\chi^{2}({\jpsi\ar\gamma\ppkk})<\chi^{2}({\jpsi\ar\gamma\gamma\ppkk})$
and $P^{2}_{t\gamma}<0.003(\rm~GeV/c)^{2}$, are used.  $P_{t\gamma}$ is
the transverse momentum of the $\ppkk$ system with respect to the
photon.  Furthermore, the requirement
$|U_{miss}|=|E_{miss}- cP_{miss}|<0.1$ $GeV/c^2$ is used to reject events with
multiple photons; here, $E_{miss}$ and $P_{miss}$ are, respectively,
the missing energy and missing momentum calculated using only the
charged particles, which are assumed to be $\ppkk$.
$\chi^{2}({\jpsi\ar\ppkk})> 20$ is required to remove $\jpsi\ar\ppkk$
background.

There are other possible backgrounds such as $\jpsi\ar \omega\kk$, $\jpsi\ar \gamma K^0_s K^{\pm} \pi^{\mp}$, etc.
The $\jpsi\ar \omega\kk$ background is suppressed with the requirement
$|M_{\ppp}-M_{\omega}|>40{\rm~MeV}/c^2$, where the $\pio$ in $\pio\pp\kk$ is
associated to the missing momentum and energy, determined using only
the charged tracks. To remove the background from $\jpsi\ar \gamma K^0_s K^{\pm}\pi^{\mp}$, events are rejected if $\chi^{2}(\jpsi\ar \gamma
K^{\pm}\pi^{\mp}\pip\pin)<\chi^{2}(\jpsi\ar \gamma\ppkk)$ when
$|M_{\pp}-M_{\ks}|<25{\rm~MeV}/c^2$. To remove backgrounds containing $\phi$,
  $|M_{\kk}-M_{\phi}|>0.02$ GeV/$c^2$ is required.

\begin{figure}[htbp!]
\begin{center}
\epsfxsize=6cm\epsffile{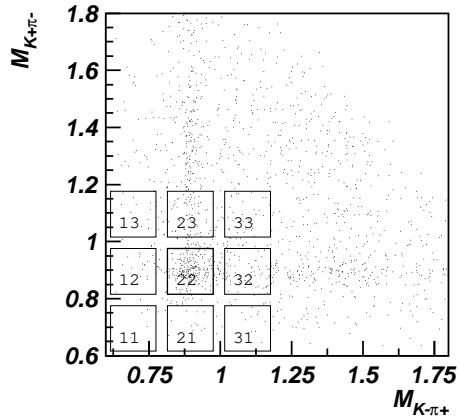}
\vspace*{5pt}
\caption{ Scatter plot of $M_{K^+ \pi^-}$ versus
  $M_{K^-\pi^+}$. Signal and sideband boxes used in the analysis are
  shown in the scatter plot.}
\label{mkstkst_box}
\end{center}
\end{figure}

A $\kstokstob$ signal is seen in the scatter plot of $M_{K^+ \pi^-}$
  versus $M_{K^-\pi^+}$, shown in Fig. \ref{mkstkst_box}. The same
  method is used as for the $\etac\ar\rr$ analysis to estimate
  sideband background.  The signal region is shown as a square box at
  (0.896,0.896) GeV/c$^{2}$ with a width of 160 MeV/c$^{2}$. The main
  backgrounds are $\etac\ar\kst K\pi$ and $\etac\ar\kk\pp$. The
  $\etac\ar\kst K\pi$ background appears as the horizontal and
  vertical bands at $m_{\kst}$, and the $\etac\ar\kk\pp$ background as
  the uniform background in the $M_{K^+ \pi^-}$ versus
  $M_{K^-\pi^+}$ scatter plot of Fig. \ref{mkstkst_box}. The
  backgrounds are estimated from the sideband boxes, which are taken
  40 MeV/c$^{2}$ away from the signal box and shown as four
  horizontal/vertical boxes (denoted as 12,21,32,23) and four diagonal
  boxes (denoted as 31,13,11,33) in Fig. \ref{mkstkst_box}. The
  horizontal and vertical sideband boxes (12,21,32,23) allow the
  determination of the the backgrounds from the horizontal and
  vertical bands, while the diagonal boxes (13,31,11,33) allow the
  estimation of the smaller uniform background contribution.
  According to a Monte Carlo study, other $\etac$ backgrounds such as
  $\etac\ar\kkkk$ and $\etac\ar\pppp$ do not survive the selection
  criteria.

\begin{figure}[htbp!]
\begin{center}
\epsfxsize=6cm\epsffile{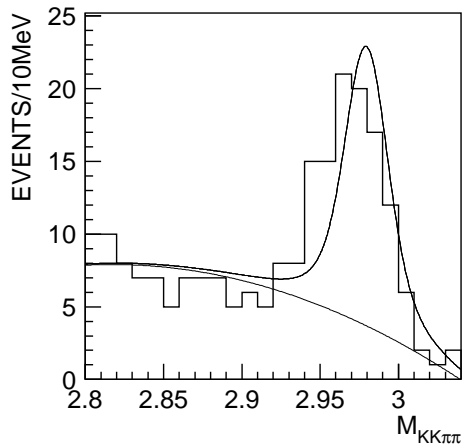}
\vspace*{5pt}
\caption{Fitting $\etac$ in the $\mppkk$
spectrum of events from signal box 22. The fitted number of $\etac$
events is $N^{sig}=80.3\pm13.6$.}
\label{netac_kk_data_a}
\end{center}
\end{figure}


\begin{figure}[htbp!]
\begin{center}
\epsfxsize=6cm\epsffile{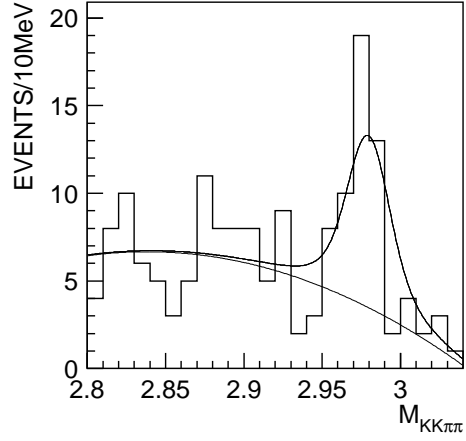}
\vspace*{5pt}
\caption{Fitting $\etac$ in the $\mppkk$ spectrum of events from the
four horizontal and vertical sideband boxes. The fitted number of
$\etac$ events is $N^{sid1}=41.0\pm10.2$.}
\label{netac_kk_data_b}
\end{center}
\end{figure}

\begin{figure}[htbp!]
\begin{center}
\epsfxsize=6cm\epsffile{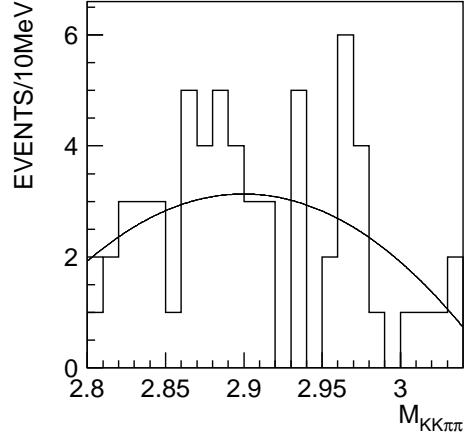}
\vspace*{5pt}
\caption{Fitting $\etac$ in the $\mppkk$ spectrum of events from the
four diagonal boxes (31,13,11,33). The fitted number of $\etac$ events
is $N^{sid2}=0.0\pm0.1$.}
\label{netac_kk_data_c}
\end{center}
\end{figure}


\subsection{Results}
Shown in Figs. \ref{netac_kk_data_a}, \ref{netac_kk_data_b}, and
\ref{netac_kk_data_c} are the $\mppkk$ fitting results for
$\kstokstob$ signal and sideband regions.  Binned maximum likelihood
fits are performed with a Breit-Wigner convoluted with a Gaussian
resolution function for the $\etac$ and a second order polynomial for
background.  The number of $\etac \ar \kstokstob$ events are obtained
from $N^{obs}=N^{sig}-N^{sid1}/2+N^{sid2}$/4. Here $N^{sig}$,
$N^{sid1}$ and $N^{sid2}$ are the fitted number of $\etac$ events in the
$\mppkk$ signal and sideband boxes(See Figs. \ref{netac_kk_data_a},
\ref{netac_kk_data_b}, \ref{netac_kk_data_c}). So
$N^{obs}=(80.3\pm13.6)-(41.0\pm10.2)/2+(0.0\pm0.1)/4=59.8\pm14.5$.

The branching fraction of $\etac \ar \kstokstob$ is given by
\begin{align}
Br(\etac \ar \kstokstob) & = \frac{N^{obs}/\varepsilon}{N_{\jpsi}\cdot
Br(\jpsi \ar \gamma \etac) \cdot Br^{2}(K^{*0}({\overline K}^{*0})\ar
K^\pm\pi^\mp)} \nonumber \\ & = (5.2 \pm 1.3) \times 10^{-3},
\nonumber
\end{align}
where $\varepsilon=3.4\%$ is the efficiency, corrected for sideband
 subtraction, determined from Monte Carlo simulation. Using
 $Br(K^{*0}({\overline K}^{*0})\ar K^\pm\pi^\mp)=0.67$ and correcting
 for charged decay modes, we obtain $Br(\etac \ar \kstkstb) = (10.4 \pm
 2.6) \times 10^{-3}$.


\section{\boldmath $\etac \ar \ww$}

  $\jpsi\ar\gamma\etac \ar \gamma\ww \ar \gamma\pppppp$ is studied to
  obtain the branching fraction of $\etac \ar \ww$.

\subsection{Event Selection}

  The charged track selection criteria and particle
  identification are the same as in the above two analyses.  Here, four
tracks must be identified as pions. The
  photon selection criteria are different for $\etac \ar
  \ww$ to get higher efficiency because there are five
  photons in this channel.
To reduce the number of spurious low energy photons produced by
secondary hadronic interactions, photon candidates must have a minimum
energy of 50 MeV and be outside a cone with a half-angle of
$6^{\circ}$ around any charged track. At least five good photons are required.

To get improved momentum resolution and to remove backgrounds, events
are kinematically fitted to the $\jpsi \ar 5\gamma\pppp$ (4C fit) and $\jpsi
\ar \gamma\pio\pio\pppp$ (6C fit) hypotheses, for all combinations of photon
candidates. $\chi^{2}_{4C}(\jpsi \ar 5\gamma\pppp)<20$ and
$\chi^{2}_{6C}(\jpsi \ar \gamma \pio \pio \pppp)< 8 $ are required. To
get cleaner $\jpsi\ar\gamma\ww $ candidates, $|M_{\pi^0}^1 -
0.135|<0.035$ GeV/$c^2$ and $|M_{\pi^0}^2 - 0.135|<0.035$ GeV/$c^2$ are
required.  Here, the photons used for $M_{\pi^0}^1$ and $M_{\pi^0}^2$
are determined from the best 6C fit to the $\jpsi \ar
\gamma\pio\pio\pppp$ hypothesis.

\subsection{Results}

\begin{figure}[htbp]
\centerline{ \hbox{\psfig{file=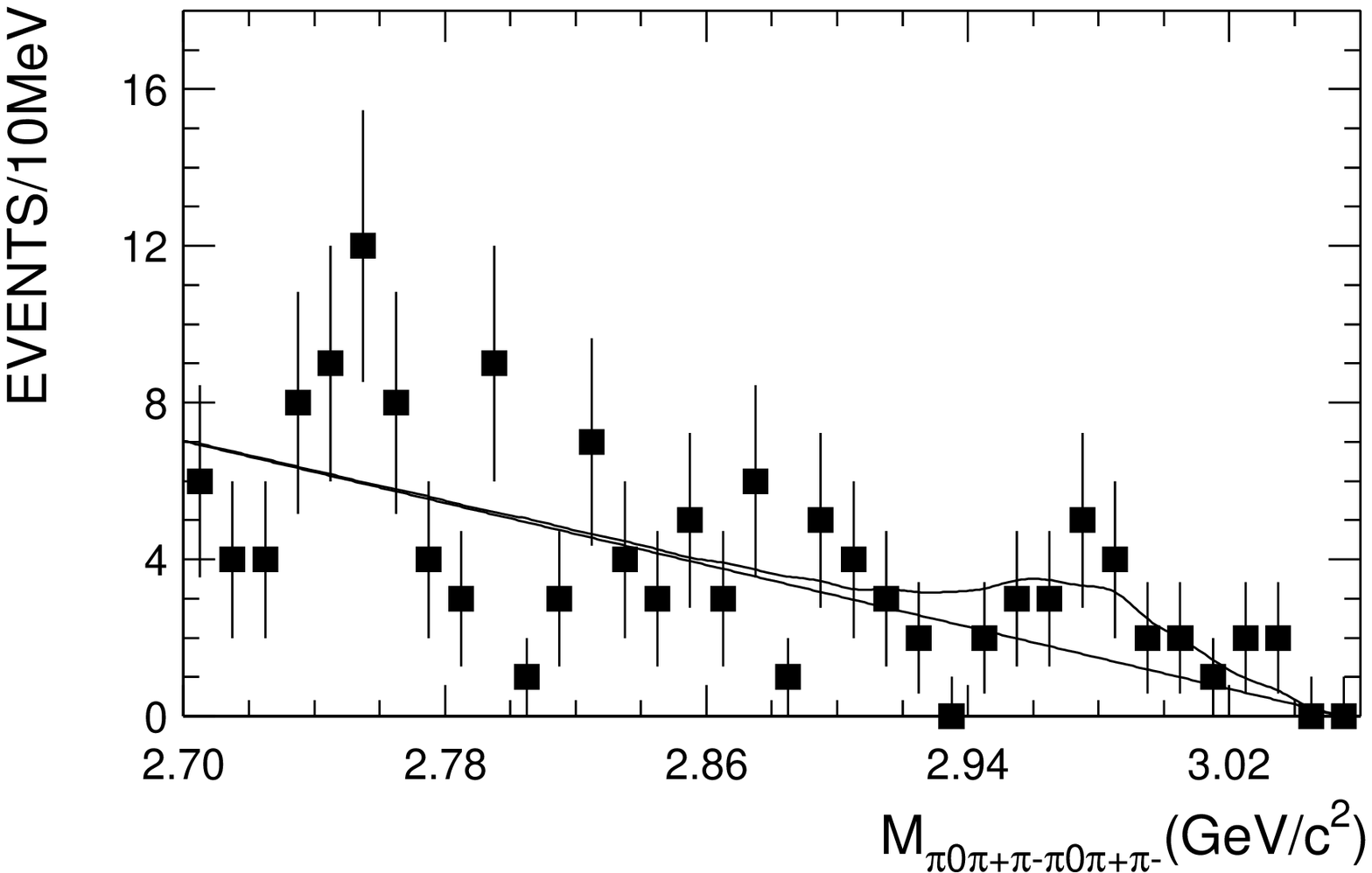, width=7cm} }}
\caption{ The $\etac$ fit of the $M_{\pppppp}$ distribution of candidate
  events in the $\ww$ region. } \label{etacmww}
\end{figure}

There are four pairs of $M_{\ppp}^1$ versus
  $M_{\ppp}^2$ combinations for every event. If one of the
  combinations satisfies $|M_{\ppp}^1-M_{\omega}|<0.035$ GeV/$c^2$ and
  $|M_{\ppp}^2-M_{\omega}|<0.035$ GeV/$c^2$, the event is selected as
  a $\jpsi \ar \gamma \ww$ candidate. The $M_{\pppppp}$
  distribution for $\jpsi \ar \gamma \ww$ candidates is shown in
  Fig. \ref{etacmww}.

 Figure \ref{etacmww} shows the result of fitting $\etac$ to the
  $M_{\pppppp}$ distribution of candidate events in the $\ww$ region. The
  shape of the $\etac$ used in the fit is obtained
  from Monte Carlo simulation of $\etac \ar \ww$. The fitted number of
  $\etac$ events is $16.4\pm 9.4$.  The result
  has large errors and includes decays other than
  $\jpsi\ar\gamma\ww$ decay.
  Therefore only the upper limit on $Br(\etac \ar \ww)$
  is determined.  At the 90\% confidence level, the upper limit on
  $N_{\etac}$ is $N^{upper}=23$.  The upper limit on Br($\etac \ar
  \ww$) at the 90\% C.L. is determined from
\begin{align}
 Br(\etac \ar \ww) & < \frac{N^{upper}/\varepsilon}{N_{\jpsi} \cdot Br(\jpsi \ar
  \gamma \etac) \cdot Br^{2}(\omega\ar\ppp) } \nonumber \\
 & = 4.0 \times 10^{-3}, \nonumber
\end{align}
where $\varepsilon=0.96\%$ is the efficiency determined from Monte
Carlo and
$Br(\omega\ar\ppp)=0.89$ \cite{pdg}.


\section{\boldmath $\etac\ar\phiphi$}

\subsection{Event Selection}

$\jpsi \ar \gamma \phiphi \ar \gkkkk$ is used to study
   $\etac\ar\phiphi$. The branching fraction of $\etac \ar \phiphi$ in
   the PDG is mainly determined from the BES result $Br(\etac \ar
   \phiphi)=(2.5 \pm 0.5 \pm 0.9) \times 10^{-3}$~\cite{dongly}. In
   order to do a
   systematic study of $\etac\ar VV$, we also analyze the $\etac \ar
   \phiphi$ decay, in order to remove the common systematic errors
   when calculating the ratios of reduced branching fractions
   $BR_R(\etac\ar VV)$.

The selection criteria for charged tracks and particle identification
  are the same as those in the $\etac\ar\rr$ analysis. At least three
  tracks must be identified as kaons. A 1C fit with a missing photon is
  performed to get better momentum resolution. To select $\jpsi \ar
  \gphiphi$ candidates, $|M_{\kk}-M_{\phi}|<20$ MeV/c$^2$, is required for
  both $\kk$ invariant masses.

\begin{figure}[htbp]
\centerline{ \hbox{\psfig{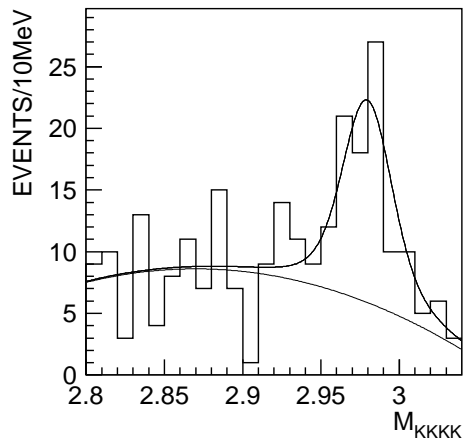} }}
\caption{ The $\etac$ fit in the $M_{\phiphi}$ distribution.  } \label{etacpp}
\end{figure}

\subsection{Results }

Fig. \ref{etacpp} shows the $\etac$ fit to the $M_{\phiphi}$
distribution. The branching fraction of $\etac \ar \phiphi$ is
determined from
\begin{align}
Br(\etac \ar \phiphi) & =
\frac{N^{obs}/\varepsilon}{N_{\jpsi}\cdot Br(\jpsi \ar
  \gamma \etac) \cdot Br^{2}(\phi \ar K^+K^-)} \nonumber \\
 & = (2.5 \pm 0.5) \times 10^{-3}, \nonumber
\end{align}
where $N^{obs}=76.1$ is the fitted number of $\etac$ events in the
$M_{\phiphi}$ distribution shown in Fig. \ref{etacpp}, $\varepsilon = 15\%$
is the efficiency determined from Monte Carlo simulation, and $Br(\phi \ar
K^+K^-)=0.491$ \cite{pdg}. This
result is consistent with the previously published BES result.

\section{\boldmath $\etac\ar \omega\phi$}

 $\jpsi\ar\gamma\etac \ar\gamma\omega\phi \ar \gamma\ppp \kk$ is
  studied to obtain the  $\etac\ar \omega\phi $ branching fraction.

\subsection{Event Selection}

 The charged track selection criteria and particle
  identification are the same as in the $\etac\ar\rr$ analysis. At least
two tracks must be identified as kaons.
 The photon selection criteria are different for $\etac \ar
   \omega\phi$ to get higher efficiency because there are three
  photons in this channel.
To reduce the number of spurious low energy photons produced by
secondary hadronic interactions, photon candidates must have a minimum
energy of 50 MeV and be outside a cone with a half-angle of
$8^{\circ}$ around any charged track. At least three good photons are required.

To get improved momentum resolution and to remove backgrounds, events
are kinematically fitted to the $\jpsi \ar \gamma \gamma \gamma\ppkk$
(4C fit) and $\jpsi \ar \gamma \pio \ppkk$ (5C fit) hypotheses,
using all photons and $\ppkk$ combinations.  The combined
probability is determined from particle identification probabilities
of all tracks and the 4C fit probability to the $\jpsi \ar \gamma
\gamma \gamma\ppkk$ hypothesis.  Candidate events must satisfy
$\chi^{2}_{4C}(\jpsi \ar \gamma \gamma \gamma \ppkk)<25$,
$\chi^{2}_{5C}(\jpsi \ar \gamma \pio \ppkk)< 20$,
$|M_{\ppp}-M_{\omega}|<0.03$ GeV/$c^2$, and $|M_{\kk}-M_{\phi}|< 0.01$
GeV/$c^2$.

\begin{figure}[htbp]
\centerline{ \hbox{\psfig{file=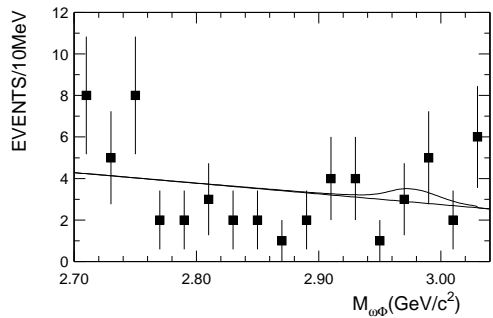, width=7cm} }}
\caption{ The $\etac$ fit to the $M_{\ppp\kk}$ distribution for
  candidate events.} \label{netac_op}
\end{figure}

\subsection{Results}


Figure \ref{netac_op} shows the result of fitting $\etac$ in the
  $M_{\ppp\kk}$ distribution of $\jpsi \ar \gamma \omega \phi$
  candidate events. The shape of the $\etac$ used in the fit is obtained
  from $\etac \ar \omega\phi$ Monte Carlo
  simulation.  The fitted number of $\etac$ events is $2.1\pm 6.1$, which has large errors and includes decays
  other than $\jpsi\ar\gamma\omega\phi$ decay.
  So only an upper
  limit on $Br(\etac \ar \omega\phi)$ is determined. At the 90\%
  confidence level, the upper limit on $N_{\etac}$ is $N^{upper}=10$.
 The upper limit on $Br(\etac \ar \omega\phi)$ at the 90\% C.L.
is determined from
\begin{align}
Br(\etac \ar \omega\phi) & < \frac{N^{upper}/\varepsilon}{N_{\jpsi}\cdot Br(\jpsi \ar \gamma \etac) \cdot Br(\phi \ar K^+K^-) \cdot Br(\omega \ar\ppp )} \nonumber \\
 & = 7.1 \times 10^{-4}, \nonumber
\end{align}
where $\varepsilon = 4.2\%$ is the efficiency determined from Monte Carlo.


\section{Systematic Errors}

Many sources of systematic errors are considered. Systematic errors
associated with efficiencies, such as from MDC tracking, particle
identification, photon selection, and kinematic fitting, are
determined by comparing $\psi(2S)$ and $\jpsi$ data with Monte Carlo
simulation for very clean decay channels, such as $\psi(2S) \ar
\pi^+\pi^-\jpsi$.

The MDC tracking efficiency has been measured using channels like
$\jpsi\ar \Lambda\overline{\Lambda}$ and $\psi(2S) \ar
\pi^+\pi^-\jpsi$, $\jpsi \ar \mu^{+} \mu^{+}$. It is found that the
efficiency of the Monte Carlo simulation agrees with that of data
within 1-2\% per charged track. The total systematic error from the
uncertainty of MDC tracking efficiency in our analysis is taken as
8\%. The particle identification (PID) efficiency systematic error is
calculated by comparing the efficiency of data with that of the Monte
Carlo. According to the study of photon detection efficiency in
$\jpsi\ar \rho\pi$ ~{\cite{lismfangss}}, SIMBES simulates the photon
detection efficiency in the full energy range within 1 to 3\%. Here,
2\% is taken as the systematic error in the detection efficiency for each
photon.

The systematic errors associated with the kinematic fit are caused by
differences between data and simulated data in the momenta and the
error matrices of charged tracks and the energies and the directions
of neutral tracks.  To check the consistency between data and Monte
Carlo simulation, two channels, $\jpsi\ar \rho^{0}\pi^{0}$ and
$\jpsi\ar \Lambda\overline{\Lambda}$, are analyzed for 2-prong and
4-prong events, respectively. The systematic error for 4-prong events
caused by the kinematic fit is determined to be 4\%. For 4-prong
events containing more than one photon, such as $\jpsi \ar \gamma
\etac\ar\gamma\ww$ and $\jpsi \ar \gamma \etac\ar\gamma\omega\phi$,
10\% is estimated as the systematic error caused by the kinematic fit.

The choice of different sideband regions can cause differences in
$Br(\etac \ar VV)$.  These differences are regarded as the systematic
error associated with the choice of sidebands.  Changing the fit
range, binning, and background shape causes some differences in the
result. The largest difference in $Br(\etac \ar VV)$ caused by these
changes is taken as the systematic error from these sources.

The $\etac$ mass resolutions used in the fits to determine the number
of $\etac$ events are determined from Monte Carlo studies.
Differences in $\etac$ mass resolutions between data
and Monte Carlo also bring uncertainty
to the determination of the branching fraction.
The differences in $Br(\etac \ar VV)$ caused by the uncertainty
of $\etac$ mass resolution are taken as the systematic error caused by
$\etac$ mass resolution.

Many selection requirements are used to remove backgrounds. To
 estimate the systematic error caused by these, different selection
 requirements are used. The differences of $Br(\etac \ar
 VV)$ caused by the change of requirements are taken as systematic errors.

The number of $\jpsi$
events is $(57.7\pm 2.7)\times 10^{6}$ determined from 4-prong events
~\cite{fangss}. The uncertainty of the number of $\jpsi$
events is 4.7\%.  The branching fraction of $\jpsi\ar\gamma\etac$ is
$(1.3\pm0.4)\%$, according to the PDG~\cite{pdg}. This will contribute 30\% to
the systematic error of $\etac\ar VV$.

\begin{table}[htpb]
\caption{Summary of systematic errors for $\etac\ar VV$ (\%).}
\begin{center}
\begin{tabular}{|l|c|c|c|c|c|}
  \hline
  Sources   & $\etac\ar\rr$     & $\etac\ar\kstkstb$ & $\etac\ar\ww$  & $\etac\ar\phiphi$  &  $\etac\ar\omega\phi$  \\ \hline
  MDC tracking efficiency                 & 8   & 8   & 8   & 8   & 8  \\
  Particle ID                             & 3   & 3   & 3   & 3   & 3  \\
  Detective efficiency of photon          & 2   & 2   & 10  & -   & 6  \\
  kinematic fit                           & 4   & 4   & 10  & 4   & 10 \\
  Different side-bands                    & 16  & 12  & -   & -   & -  \\
  Fit range, binning and background shape & 3   & 7   & -   & 11  & -  \\
  $\etac$ mass resolution                 & 7   & 8   & -   & 7   & -  \\
  Different Cuts                          & 18  & 19  & 14  & 11  & 32 \\
  The number of $\jpsi$ events            & 4.7 & 4.7 & 4.7 & 4.7 & 4.7\\
  $Br(\jpsi\ar\gamma\etac) $              & 30  & 30  & 30  & 30  & 30 \\\hline
  Total                                   & 40.7& 41  & 37  & 36  & 46 \\ \hline
\end{tabular}
\label{syserrrr}
\end{center}
\end{table}

Table \ref{syserrrr} lists all systematic error contributions and the total
systematic errors.
The $Br(\etac\ar VV)$ and $Br(\jpsi\ar\gamma\etac)
\cdot Br(\etac\ar VV)$ results including systematic errors are listed
in Tables \ref{cmpbespdg} and \ref{jpsitogetac}, respectively. Table
\ref{jpsitogetac} results do not have the large systematic error
from $Br(\jpsi\ar\gamma\etac)$.

\begin{table}[htpb]
\caption{Comparison of $Br(\etac\ar VV)$ values between BES and the PDG ~\cite{pdg}.}
\begin{center}
\begin{tabular}{|l|c|c|}
  \hline
  Br  & BES & PDG\\ \hline
  $Br(\etac \ar \rho \rho)$ & $(1.25 \pm 0.37 \pm 0.51) \times 10^{-2}$ & $(2.6 \pm 0.9 ) \times 10^{-2}$ \\
  $Br(\etac \ar \kstkstb)$ &  $ (10.4 \pm 2.6\pm 4.3)  \times 10^{-3}$ & $(8.5 \pm 3.1 ) \times 10^{-3}$ \\
  $ Br(\etac \ar \ww)$ &  $<  6.3 \times 10^{-3} $ & $<  3.1 \times 10^{-3}$ \\
  $Br(\etac \ar \phiphi)$ &  $(2.5 \pm 0.5 \pm 0.9) \times 10^{-3}$ & $(2.6 \pm 0.9) \times 10^{-3}$ \\
  $Br(\etac \ar \omega\phi)$ &  $<  1.3 \times 10^{-3} $ & - \\\hline
\end{tabular}
\label{cmpbespdg}
\end{center}
\end{table}

\begin{table}[htpb]
\caption{Summary of $Br(\jpsi\ar\gamma\etac) \cdot Br(\etac\ar VV)$. }
\begin{center}
\begin{tabular}{|l|c|}
  \hline
  Br  & BES \\ \hline
  $Br(\jpsi\ar\gamma\etac) \cdot Br(\etac \ar \rho \rho)$ & $(1.6 \pm 0.6 \pm 0.4) \times 10^{-4}$\\
  $Br(\jpsi\ar\gamma\etac) \cdot Br(\etac \ar \kstkstb)$ &  $( 1.4 \pm 0.3 \pm 0.5 ) \times
 10^{-4}$ \\
  $Br(\jpsi\ar\gamma\etac) \cdot  Br(\etac \ar \ww)$ & $<  6.7 \times 10^{-5}$ \\
  $Br(\jpsi\ar\gamma\etac) \cdot Br(\etac \ar \phiphi)$ & $(3.3 \pm 0.6 \pm 0.6 ) \times
 10^{-5}$ \\
  $Br(\jpsi\ar\gamma\etac) \cdot Br(\etac \ar \omega\phi)$ & $<  1.4 \times 10^{-5} $ \\\hline
\end{tabular}
\label{jpsitogetac}
\end{center}
\end{table}


\section{Summary}

The $Br(\etac\ar VV)$ results are compared with PDG values in the
Table \ref{cmpbespdg}. Table \ref{tablebrcmp} compares the ratio
of branching fractions with expectations from SU(3) symmetry. The
ratios are more consistent with SU(3) symmetry than ratios based
on PDG branching fractions, as shown in Table \ref{tablebr}. Our
$Br(\etac \ar \rho \rho)$ result is consistent with the model
prediction for $\Gamma(\etac\ar\rr)$ of Ref.~\cite{etacvv_width}.
The upper limit on $ Br(\etac \ar \omega\phi)$ is also given.


\begin{table}[htpb]
\caption{Check of SU(3) symmetry with new BES results.  The common
  systematic error has been removed. }
\begin{center}
\begin{tabular}{|l|c|c||c|c|}
  \hline
     & BR($10^{-3}$) & $BR_{R}$($10^{-3}$) & Ratio of $BR_{R}$ & SU(3) \\ \hline
  $Br(\etac\ar\rho\rho)$ & $12.5\pm4.9$ & $6.3\pm2.4$ & $3.1\pm1.2$ & 3 \\
  $Br(\etac\ar\kstkstb)$ & $10.4\pm3.2$   & $6.1\pm1.9$ & $3.0\pm1.0$ & 4  \\
  $Br(\etac\ar\ww)$      & $<6.3 $    & $<3.2$ & $<1.6 $ & 1  \\
  $Br(\etac\ar\phiphi)$ & $2.5\pm 0.7$& $2.0\pm0.6 $ & $1\pm0.3 $ & 1  \\ \hline
\end{tabular}
\label{tablebrcmp}
\end{center}
\end{table}

\section{Acknowledgment}
\vspace{0.4cm}

The BES collaboration thanks the staff of BEPC for their hard
efforts. We also thank Prof. Guangda Zhao for helpful
discussions. This work is supported in part by the National Natural
Science Foundation of China under contracts Nos. 10491300,
10225524, 10225525, 10425523, the Chinese Academy of Sciences under
contract No. KJ 95T-03, the 100 Talents Program of CAS under
Contract Nos. U-11, U-24, U-25, and the Knowledge Innovation
Project of CAS under Contract Nos. U-602, U-34 (IHEP), the
National Natural Science Foundation of China under Contract No.
10225522 (Tsinghua University), and the Department of Energy under
Contract No.DE-FG02-04ER41291 (U Hawaii).


\end{document}